\documentclass[11pt]{article}
\usepackage{amsmath,amsfonts}
\def \beq  {\begin{equation}}
\def \eeq  {\end{equation}}
\def \beqar {\begin{eqnarray}}
\def \eeqar {\end{eqnarray}}
\def\sqr#1#2{{\vcenter{\vbox{\hrule height.#2pt
\hbox{\vrule width.#2pt height#1pt \kern#1pt
\vrule width.#2pt}\hrule height.#2pt}}}}

\def\vx {{\vec x}}
\def\vy {{\vec y}}

\def\vf {{\varphi}}

\def\dag {{\dagger}}

\def\Tr {{\rm Tr}}

\def\bu {\bar{u}}

\def\vx {{\vec x}}

\def\vy{\vec{y}}

\def\dag {\dagger}
\def\del {\partial}

\def\a {\alpha}
\def\b {\beta}
\def\e {\epsilon}
\def\d {\delta}
\def\s {\sigma}
\def\l {\lambda}
\def\o {\omega}

\def\C {{\cal C}}

\def\E {{\cal E}}

\def\vf {{\varphi}}

\def \C {{\cal C}}

\def\half{\textstyle{1\over 2}}

\begin{document}
\def \CMP {{ Commun. Math. Phys.}}
\def \PRL {{ Phys. Rev. Lett.}}
\def \PL {{Phys. Lett.}}
\def \NPBProc {{ Nucl. Phys. B (Proc. Suppl.)}}
\def \NP {{ Nucl. Phys.}}
\def \RMP {{ Rev. Mod. Phys.}}
\def \JGP {{ J. Geom. Phys.}}
\def \CQG {{ Class. Quant. Grav.}}
\def \MPL {{Mod. Phys. Lett.}}
\def \IJMP {{ Int. J. Mod. Phys.}}
\def \JHEP {{ JHEP}}
\def \PR {{Phys. Rev.}}
\def \JMP {{J. Math. Phys.}}
\def\JoP {{J. Phys.}}
\begin{titlepage}
\null\vspace{-62pt} \pagestyle{empty}
\begin{center}
\rightline{CCNY-HEP-06/7}
\rightline{May 2006}
\vspace{1truein} {\Large\bfseries Color Skyrmions in the Quark-Gluon Plasma}\\
\vskip .2in\noindent

\vspace{.5in}
{\bf\large JIAN DAI}\footnote{E-mail: \fontfamily{cmtt}\fontsize{11pt}{15pt}\selectfont
jdai@sci.ccny.cuny.edu} and {\bf\large V. P. NAIR}\footnote{E-mail: \fontfamily{cmtt}\fontsize{11pt}{15pt}\selectfont
vpn@sci.ccny.cuny.edu}\\
\vspace{.15in}{\itshape Physics Department\\
City College of the CUNY\\
New York, NY 10031}\\

\fontfamily{cmr}\fontsize{11pt}{15pt}\selectfont
 \vskip 1in
\centerline{\large\bf Abstract}
\end{center}
We consider the general formulation of nonabelian fluid dynamics based on symmetry considerations. We point out that, quite generally, this admits solitonic excitations
which are the color analog of skyrmions. Some general properties of the solitons are discussed.

\end{titlepage}
\pagestyle{plain} \setcounter{page}{2}
\setcounter{footnote}{0}

\section{Introduction}
 The collision of heavy nuclei at the Relativistic Heavy Ion Collider has created an interesting
 new state of matter \cite{expt}. It is clear that a state with
 deconfinement of quarks and gluons has been achieved. There are
 various indications that the resulting fluid is probably best
 described as a color liquid. One of the surprises has been the very
 low shear viscosity of the fluid. The value of the viscosity, it
 seems, is close to what may be the theoretical lower limit possible,
 a value which may be understandable in terms of a gravity-dual to the theory
 \cite{grav}.
Thus a good first approximation to the description of this color
 liquid may be as a `perfect fluid'. However, unlike the case of ordinary fluids,
 since the constituents
 carry color degrees of freedom, the
transport of  such degrees of freedom, in a way
 consistent with the nonabelian gauge symmetry, becomes an important issue
 for this `color liquid'.
 This can, of course, be studied via kinetic
 equations, starting from the basics of QCD. However, the hierarchy of kinetic
 equations has to be truncated for reasons of computability, very often to the level of
 uncorrelated single-particle distribution functions,
 and such an
 approach is then limited to dilute systems near equilibrium.
Since the experimental results indicate that the color liquid is not
 a dilute system, the validity and fruitfulness of this approach become
 questionable.
By
 contrast, experience with ordinary fluid dynamics shows that one can
 derive the equations of fluid dynamics from very general principles,
 which then shows that the equations have a range of validity
 significantly beyond the regime where the truncated kinetic equations apply.
 This latter, {\it a priori}, approach has been developed for nonabelian
 fluid dynamics as well \cite{JNP, bistro}. The nonabelian degrees of freedom of the
 fluid are described by a field which takes values in the color
 group, for example, by $g(\vec{x},t)\in SU(3)$ for QCD. An immediate
 and qualitatively striking consequence is that nonabelian fluid
 dynamics leads to topological solitons. These are color skyrmions
 associated with the homotopy group $\Pi_3[SU(3)]=\mathbb{Z}$. The
 topological quantum number is the color analog of what has been
 known as helicity in ordinary fluid dynamics for a long time.
 The purpose of this article is to point out the existence of such solitons,
 or configurations of nonzero color helicity, and study some of their properties.

 In the next section, we will review briefly the formulation of
 ordinary fluid mechanics in group theory language and its
 generalization to include the transport of nonabelian degrees of
 freedom. This subject has been reviewed recently in \cite{JNPP}.
 In section 3, we will introduce the solitons and work out
 their properties. We then conclude with a short discussion.

\section{Group theory and fluid dynamics}

 We begin with the well-known observation that ordinary fluid
 dynamics can be described as a Poisson bracket system. With $\rho$ as
 the fluid density and $v_i$ as the fluid velocity, the fundamental Poisson
 bracket relations are given by
 \beqar
  [~\rho(\vx,t),\rho(\vy,t)]&=&0,\nonumber\\
  {[}v_i(\vx,t),\rho(\vy,t){]}&=&{\del\over \del
  x^i}~\d^{(3)}(x-y)\label{GT1}\\
  {[}v_i(\vx,t),v_j(\vy,t){]}&=&-{\o_{ij}\over\rho}~\d^{(3)}(x-y)\nonumber
 \eeqar
where $\omega_{ij}$ is the vorticity defined by
 \beq
  \o_{ij}=\del_iv_j-\del_jv_i\label{GT2}
 \eeq
 Equations (\ref{GT1}) will lead to the usual equations of fluid dynamics, where the
 Hamiltonian $H$ may be taken as
 \beq
  H=\int d^3x~\left[{1\over 2}~\rho~ v^2+V(\rho)\right]\label{GT3}
 \eeq
 The canonical structure, or the symplectic form, defined by the
 Lagrangian is the inverse of the fundamental Poisson bracket. In
 the present case, the fundamental Poisson brackets have a zero mode
 and so this is a complication in finding a suitable Lagrangian.
 This can be seen in terms of the Chern-Simons action for the
 velocity, namely,
 \beq
  \C={1\over 8\pi}\int \e^{ijk}v_i\del_jv_k \label{GT4}
 \eeq
 (The invariant defined by $\C$ is the fluid helicity.) It is easily
 seen that $\C$ Poisson commutes with any observable $F$,
 \beq
  [F,\C]\equiv \int ~\left[{\d F\over \d \rho}{\del \over \del x^i}\left({\d \C\over \d
  v_i}\right)-{\d \C\over \d \rho}{\del \over \del x^i}\left({\d F\over \d
  v_i}\right)-\o_{ij}{\d F\over \d v_i}{\d \C\over \d
  v_j}\right]=0 \label{GT5}
 \eeq
 This shows that $\d\C/\d v_i$ is a zero mode for the brackets, preventing the inversion needed to define the action.
 The solution to this problem is also rather clear. Since $\C$
 commutes with all observables, we must fix its value and consider
 only variations in $v_i$ which preserve this value. In particular,
 if we choose $\C=0$, we must consider $v_i$ which trivializes the
 Chern-Simons term by making $v_i\del_j v_k\e^{ijk}$ into a total
 derivative. Such $v_i$ are given in terms of the Clebsch
 parametrization,
 \beq
  v_i=\del_i\theta+\a~\del_i\b ,\label{GT6}
 \eeq
 where $\theta$, $\a$, $\b$ are arbitrary functions.

 A suitable action which leads to the Poisson bracket relations
 (\ref{GT1}), and the continuity and Euler equations, is
 \beq
  S=\int dtd^3x~\left[\rho~\dot{\theta}+\rho~\a~\dot{\b}-{1\over 2}\rho
  ~v^2-V(\rho )\right] \label{GT7}
 \eeq
 Notice that we have the canonically conjugate pairs
 $(\rho,\theta)$, $(\rho\a,\b)$.

 We will now rewrite this in terms of a group element
 \beq
  g={1\over
  \sqrt{1-u\tilde{u}}}
  \left(\begin{array}{cc}1&u\\\tilde{u}&1\end{array}\right)
  \left(\begin{array}{cc}e^{i\theta/2}&0\\0&e^{-i\theta/2}\end{array}\right)
  \label{GT8}
 \eeq
 where $u$ is a complex variable. $g$ is an element of $SU(1,1)$ if
 $\tilde{u}=\bu$ and an element of $SU(2)$ if $\tilde{u}=-\bu$. (We
 will consider both possibilities together for a while.) Using
 (\ref{GT8}), we find by direct computation that \cite{JP, JNP}
 \beq
  -i\Tr(\sigma_3 g^{-1}dg)=d\theta+\a d\b \label{GT9}
 \eeq
 where
 \beq
  \a={u\tilde{u}\over 1-u\tilde{u}}~,\hskip .2in
  \b=-{i\over 2}\log({u/\tilde{u}})\label{GT10}
 \eeq
 Equations (\ref{GT9},\ref{GT10}) show that the Clebsch
 parametrization can be expressed in terms of a single group element
 as $v_i=-i\Tr(\s_3g^{-1}\del_ig)$. Notice that $\theta$ in
 (\ref{GT8}) corresponds to the $\s_3$-direction in $g$; thus $\a$,
 $\b$ parametrize the space $SU(1,1)/U(1)$ and $SU(2)/(1)$,
 respectively, for $\tilde{u}=\bu$ and $\tilde{u}=-\bu$. Going back
 to (\ref{GT7}) we notice that the action can be written as
 \beq
  S=\int d^4x~\left[ -ij^\mu \Tr (\s_3g^{-1}\del_\mu g)-\left( {j^i j^i\over 2\rho }+V\right)\right]
  \label{GT11}
 \eeq
 where $j^0=\rho$. Elimination of $j^i$ using its equation of
 motion takes us back to (\ref{GT7}). The form of equation (\ref{GT11})
 also shows us how to generalize to a relativistic situation,
 \beq
  S=\int d^4x~ \bigl[ -ij^\mu ~\Tr(\s_3g^{-1}\del_\mu g)-F(n)\bigr]\label{GT12}
 \eeq
 where $n^2=j^\mu j^\nu g_{\mu\nu}$, $g_{\mu\nu}$ being the metric tensor. The choice of the function $F$
 specifies the equation of state for the fluid. A four-velocity
 $u^\mu$ for the fluid can be defined by $j^\mu =nu^\mu$. The
 energy-momentum tensor is then
 \beq
  T^{\mu\nu}=nF^\prime u^\mu u^\nu-g^{\mu\nu}(nF^\prime-F) \label{GT13}
 \eeq
 which also identifies the pressure as $p=nF^\prime -F$.

 The Poisson bracket relations which follow from (\ref{GT11}) (or
 (\ref{GT12})) are
 \beqar
  [\rho(\vx ,t),\rho(\vy, t)]&=&0\nonumber\\
  {[}\rho(\vx , t), ~g(\vy, t){]}&=&-ig(x)~t_3~\d^{(3)}(x-y)\label{GT14}
 \eeqar
 Here $t_a = \half \sigma_a $. This equation shows that $\rho$ generates right translations on $g$ along the
 $\s_3$-direction. In the quantum theory, it is easily checked that
 $U=\exp[-2\pi i\int\rho]$ acts on $g$ as
 \beq
  U^\dag g ~U=g~e^{2\pi it_3}=-g\label{GT15}
 \eeq
 All observables are invariant under $U$ since they have even powers
 of $g$, $g^\dag$. This means that $U$ is the identity operator on
 observables, which, in turn, implies that
 $\int\rho=\mbox{integer}$ \cite{ray}. This is essentially the statement that
 the fluid is made of particles. Mathematically, this is due to the
 $\theta$-direction being compact. Turning this logic around, we
 see that it is consistent to require that the field $\theta$, which appears in the Clebsch
 parametrization, should be a compact direction since fluids of
 interest are ultimately made of particles.

 From the Clebsch parametrization,
 $\o_{ij}=(\del_i\a\del_j\b-\del_j\a\del_i\b)$. Consider now the
 integral of this over some compact region $V$ in $\mathbb{R}^2$.
 $\int_V\o$ is the volume of $SU(1,1)/U(1)$ or $SU(2)/U(1)$ over the
 image of $V$ via the map $g(x)$. This will be quantized for
 $S^2=SU(2)/U(1)$, but not for $SU(1,1)/U(1)$. Thus we see that the choice of the
 group is determined by the quantization of $\int \rho$ and by
 whether we want quantized vorticity or not.

 We now turn to the generalization of this to other groups. A
 Clebsch-type parametrization for higher groups would be an obvious
 direction to try \cite{JNP}. This is possible for the groups $SO(n)$ with
 $v_i^a=-i\Tr(t^aR^T\del_iR)$ where $R\in SO(2n-1)$ and $t_a$ are
 generators of $SO(n)\subset SO(2n-1)$. The relevant coset is
 $SO(2n-1)/SO(n)\times SO(n-1)$. The resulting dynamics is rather
 involved and does not admit the Eckart factorization $J_i^a=Q^aj_i$
 which is what we would expect if the nonabelian charge density
 $Q^a$ is transported by particle motion.

 A different approach is to start with the motion of particles
 carrying nonabelian charges. This is described by the Wong
 equations \cite{wong}
 \beqar
  \dot{Q}^a-f^{abc}A^c_i ~\dot{x}^i ~Q^b&=&0\nonumber\\
  \dot{p}_i-F^a_{ij}~\dot{x}^jQ^a&=&0\label{GT16}
 \eeqar
 These equations of motion can be obtained from the action \cite{bal}
 \beq
  S=\int dt~\left[{1\over 2}m\dot{x}^2+A_i^aQ^a\dot{x}^i\right]-in\int dt~
  \Tr(\s_3g^{-1} \dot{g})\label{GT17}
 \eeq
 where $Q^a= \Tr(\s_3g^{-1} t^a g)$ and $g\in SU(2)$. (Here we are
 considering $SU(2)$ color for simplicity. We will extend this to
 any Lie group shortly.) The first term is the usual action for a
 particle coupled to an external gauge field except for the color
 charge factor $Q^a$. The second term is what leads to the dynamics
 of the color charge. Notice that $Q^a$ is invariant under
 $g\rightarrow gh$, where $h=\exp(i t_3 \vf)$. The second
 term in the action changes by a surface term $\Delta
 S=n\int\dot{\vf}dt=n\Delta\vf$. Thus, the equations of motion are
 invariant under $g\rightarrow gh$. However, one can consider closed
 loops in the $U(1)$-subgroup, generated as a trajectory over time,
 which would have $\Delta\vf=2\pi$. The
 invariance of $e^{iS}$ then requires $n\in \mathbb{Z}$. For integer
 values of $n$, the theory (including the quantum theory) involves
 only $SU(2)/U(1)=S^2$ variables. In fact, the second term of $S$
 gives the symplectic form $\Omega=in\Tr(\s_3g^{-1}dg\wedge g^{-1}dg)$
 for the parameters in $g$. Since the phase space for this is the
 two-sphere, the phase volume is finite, giving a finite-dimensional
 Hilbert space upon quantization. In fact, the quantization of
 (\ref{GT17}) leads to one unitary irreducible representation of
 $SU(2)$ with a highest weight state of $j=n/2$. The action
 (\ref{GT17}) thus leads to the standard description of color by
 matrices. The term $-in\Tr(\s_3g^{-1}dg)$ is often referred to as
 the Kostant-Kirillov-Souriau (KKS) form \cite{nair}.

 Focusing on the color degrees of freedom, the generalization of
 (the second term of) the action (\ref{GT17}) to many particles is given by
 \beq
  S=-i\int dt \sum\limits_\l ~n_\l ~\Tr(\s_3 ~g_\l^{-1}\dot{g}_\l)\label{GT18}
 \eeq
 where $\l=1,2,\ldots, N$, labels the particle under consideration.
 We can now take the continuum limit by taking $N$ large, with an
 average volume $v$ per particle. This means that we can write
 $\l\rightarrow \vx$, $\sum_\l\rightarrow \int d^3x/v$ and
 $n_\l/v=j_0$. The action (\ref{GT18}) then becomes
 \beq
  S=-i\int d^4x~ j^0~\Tr(\s_3g^{-1}\dot{g})\label{GT19}
 \eeq
 where $g=g(\vx,t)$. Notice the similarity of this to the first term
 of the action for ordinary fluid dynamics. Taking (\ref{GT19}) as the
 key term which leads to the symplectic form for the color degrees of
 freedom, we can write the
 action for fluids, where the particle carry $SU(2)$ color degrees
 of freedom, as
 \beq
  S=\int d^4x~ \left[-i j^\mu~\Tr(\s_3 g^{-1} D_\mu g)-F(n)-{1\over
  4}F^a_{\mu\nu}F^{a\mu\nu}\right]\label{GT20}
 \eeq
 where $D_\mu g=\del_\mu g+A_\mu g$ and $A_\mu=-i t^a A^a_\mu$ is
 a nonabelian background field. (By this we mean the field background in which the fluid moves; thus $A_\mu$ represents gluon degrees of 
 freedom which are not homogenized with the fluid, such as hard gluons.)
  $j^\mu$ is a current
 four-vector, $n^2=j^\mu j^\nu g_{\mu\nu}$. $j^i$ can be eliminated
 by its equation of motion to get a simpler form of $S$, which,
 however, will not be manifestly Lorentz invariant. The dynamics which
 follows from (\ref{GT20}) will be the $SU(2)$ analog of
 magnetohydrodynamics. The color current, which couples to
 $A^a_\mu$, is easily seen to be
 \beq
  J^{a\mu}= \Tr(\s_3g^{-1} t^a g) ~j^\mu=Q^aj^\mu \label{GT21}
 \eeq
 We see that the Eckart factorization is realized
 with $Q^a$ as the charge density for the fluid. The equations of
 motion are given as
 \beqar
D_\mu J^{a\mu} &=&0\nonumber\\
  F^\prime u^\mu &=& -i\Tr(\s_3g^{-1}D_\mu g)\label{GT22}
 \eeqar
 where the color flow velocity $u^\mu$ is defined by
 $j^\mu=nu^\mu$.  We can also carry out a variation on $g$ on the right, which leads to the
 equation $\del_\mu j^\mu =0$. This equation is not independent of equations
 (\ref{GT22}).
 The
 energy-momentum tensor for the fluid is given by
 \beq
  T^{\mu\nu}=nF^\prime u^\mu u^\nu-g^{\mu\nu}(nF^\prime-F)\label{GT23}
 \eeq
 and obeys the expected relation
 \beq
  \del_\mu T^{\mu\nu}=\Tr(J_\mu F^{\mu\nu})\label{GT24}
 \eeq

 It is interesting at this stage to give an interpretation of the
 group element $g$. Let $\rho=\rho^at^a$ be the
 nonabelian charge density of a distribution of particles. Under a
 gauge transformation $U\in SU(2)$, it transforms as
 \beq
  \rho\rightarrow \rho^\prime = U^{-1}\rho U\label{GT25}
 \eeq
 We can diagonalize the hermitian matrix $\rho$ at each point by an
 $ (\vx,t)$-dependent transformation. We may thus write
 $\rho=g~\rho_{diag}~g^{-1}$, or
 \beq
  \rho^a=\rho_0~\Tr(\s_3g^{-1}t^a g)=n~\Tr ( \s_3 g^{-1}t^a g)\label{GT26}
 \eeq
 where $\rho_0=n$ is the eigenvalue of $\rho$. Evidently, $n$ is
 gauge invariant. Comparing with our previous expression for
 $\rho^a=J^{a0}$, we see that the dynamical variable $g(\vx,t)$ can
 be interpreted as the gauge transformation which diagonalizes the
 charge density at each point. The flow of the gauge invariant
 eigenvalues is given by $u^\mu$. Under a gauge transformation,
 $g\rightarrow U^{-1}g$.

 The generalization to higher groups is clear from this discussion.
 For a group $G$, we have $rank(G)$ eigenvalues and hence $rank(G)$
 $n$'s and $u^\mu$'s. The field $g$ is an element of $G$ and the
 action is given by
 \beq
  S=-i\sum\limits_s^{rank(G)}\int d^4x~ j_s^\mu ~\Tr(q_s ~g^{-1}D_\mu
  g)- \int d^4x~F(n_1,n_2,\ldots) ~+~ S_{YM}\label{GT27}
 \eeq
 $q_s$ are the diagonal generators of $G$, $j_s^2=n^2_s$. We have as
 many $n$'s as there are simultaneously diagonalizable conserved
 charges.

 For a statistical distribution, we expect a chemical potential for
 each of the simultaneously diagonalizable conserved charges. The
 values of these chemical potentials are fixed by the values of the
 corresponding charges. Thus a statistical distribution of particles
 is specified by the values it has for the simultaneously
 diagonalizable conserved charges. What we find for the fluid is in
 conformity with this.

 One can give a general justification for (\ref{GT27}) starting from a
 one-particle action again. The relevant observation is the
 following. The general form of KKS action is \cite{nair}
 \beq
  S=-i\sum\limits_sw_s\int dt ~\Tr(q_s~g^{-1}~\dot{g})\label{GT28}
 \eeq
 There are quantization condition on this action. The numbers
 $(w_1,w_2,\ldots)$ should form the weight vector of the highest
 weight state of a unitary irreducible representation of the group.
 Upon quantization, the action (\ref{GT28}) then gives a Hilbert space
 which corresponds to the unitary irreducible representation
 characterized by $(w_1,w_2,\ldots)$. What we have given in
 (\ref{GT27}) is indeed the appropriate generalization of this result
 to fluids. (The action (\ref{GT28}) leads to the symplectic form which is
 the K\"ahler two-form on the coadjoint orbit of the element
 $\sum_s w_s q_s$. So this method is also known as the coadjoint orbit
 method.)

 The foregoing discussion also shows that ordinary hydrodynamics is
 a special case of this general structure where we have only one
 conserved charge, namely, the particle number.

 For the sake of completeness, we give the equations of motion for
 the action (\ref{GT27}); these are
 \beqar
  \del_\mu j^\mu_s&=&0\nonumber\\
   \sum\limits_sj^\mu_s(D_\mu Q_s)^a&=&0\label{GT29}\\
  u^\mu_s~{\del F\over\del n_s} &=& -i \Tr(q_s~g^{-1}D_\mu g)\nonumber
 \eeqar
 where $Q_s^a = \Tr (q_s g^{-1}t^a g)$.
 The basic Poisson brackets are given by
 \beqar
  [\rho^a(\vx, t),\rho^b(\vy, t)]&=&f^{abc}\rho^c(\vx, t)~\d^{(3)}(x-y)\nonumber\\
  {[}j^0_s(\vx, t), j^0_{s^\prime}(\vy, t){]}&=&0\label{GT30}\\
  {[}~j^0_s(\vx, t), g(\vy, t){]}&=&-i g(x)~q_s~\d^{(3)}(x-y)\nonumber
 \eeqar
Here $\rho^a=J^{a0}$ is the color charge density and
the color current is $J^{a\mu}=\sum_sj^\mu_s Q^a_s$.

\section{Color skyrmions}

 The description of nonabelian charge using the KKS form is a
 standard part of Lie group theory. In fact, if we ask whether we
 can find a classical theory which upon quantization gives
 finite-dimensional Lie algebra (color) matrices in a single
 irreducible representation, the answer is the KKS action
 (\ref{GT28}). Since the action for fluid dynamics which we postulate
 is a straightforward fluid generalization of this basic result for
 particles, we see that, quite generally, the color degrees of
 freedom of the fluid are described by $g(\vx,t)\in G$ and
 $j^0_s\in\mbox{Cartan subalgebra of $G$}$. For QCD, we have
 $g(\vx,t)\in SU(3)$. At a fixed time, we have $g(\vx): V\rightarrow SU(3)$,
 where $V$ is a region in $\mathbb{R}^3$ in which the fluid exists.
 For configurations with $g\rightarrow 1$ on the boundary of $V$,
 these functions $g(\vx)$ are equivalent
 to $g(\vx):S^3\rightarrow SU(3)$. The homotopy classes, or equivalence classes up to smooth deformations, of such maps
 are given by $\Pi_3[SU(3)]=\mathbb{Z}$. The topologically
 nontrivial configurations of $g(\vx)$ are then solitons. These are
 mathematically the same as the usual skyrmions, although these are in the
 color sector and not in the flavor sector \cite{skyrmion, witten} {\footnote{Skyrmions where the target space includes color have been considered in a different and unrelated context in \cite{kaplan}.}}. The soliton number which
 characterizes the homotopy classes $\Pi_3[SU(3)]$, or the
 skyrmions, is
 \beq
  Q=-{1\over 24\pi^2}\int d^3x~ \e^{ijk}~
  \Tr (g^{-1}\del_ig~g^{-1}\del_jg~g^{-1}\del_k g)  \label{CS1}
 \eeq
 (This is essentially the color version of helicity.)

 For $SU(3)$, there are two distinct types of maps $g:S^3\rightarrow
 SU(3)$ we can consider. One of them corresponds to the image of
 $S^3$ being an $SU(2)$ subgroup of $SU(3)$, the other corresponds
 to $S^3$ being mapped to the $SO(3)$ subgroup defined by the generators
 $(\l_2,-\l_5,\l_7)$, $\l_a$ being the usual Gell-Mann matrices of
 $SU(3)$. Most of the features of the skyrmions we are considering
 will be clear from the first type of maps, namely,
 $g:S^3\rightarrow SU(2)\subset SU(3)$, so we shall consider only
 these in this paper. This means that, effectively, we can restrict attention to
 $SU(2)\subset SU(3)$.

 The explicit solution of the equation of motion to obtain the
 soliton will be very involved and will depend sensitively on the
 choice of Hamiltonian. But, as is usually done in the case of
 flavor skyrmions, we can choose an ansatz in a given topological
 sector, which depends on some parameters, and then variationally
 minimize the energy to fix the values of these parameters. In
 general, this will give a good qualitative (and, to some extent,
 even quantitative) description of the soliton. The simplest ansatz
 we can take for $g(\vx)$ is the spherically symmetric ansatz
 \beq
  g_s(\vx)=\cos \phi (r)+i ~\s\cdot \hat{x}~\sin \phi  (r)\label{CS2}
 \eeq
 which leads to
 \beq
  Q={\phi (0)-\phi (\infty)\over \pi} \label{CS3}
 \eeq
 The profile of $\phi$ as a function of $r$ is not yet determined; for
 $Q=1$, we need one step of height $\pi$ as we go from $r=0$ to
 $r=\infty$. The simplest choice is the stereographic ansatz
 \beq
  \sin \phi = {2Rr\over R^2+r^2},~~~~
  \cos \phi = {R^2-r^2\over R^2+r^2}\label{CS4}
 \eeq
 where $R$ sets the scale for the soliton. In what follows we will
 make this simple choice.

 Up to this point we have not chosen a particular form for the
 Hamiltonian. The action (\ref{GT27}) has the form
 \beq
  S=-i\int d^4x~ j^0~\Tr(\s_3g^{-1} \dot{g})~-\int dt~H \label{CS5}
 \eeq
 $H$ is determined by the choice of $F(n)$. We will take a simple
 form which is easy to work with. More specific choices, based on
 the equation of state of the color liquid, can be made. We do not
 expect the general features of the solitons to be changed by this.
 Our choice for $F(n)$ is then
 \beq
 F(n)={1\over 2\mu^2}~n^2={1\over
 2\mu^2}\left[ j^0 j^0- \vec{j}\cdot\vec{j}\right]
 \label{CS6}
 \eeq
Eliminating $\vec{j}$ from
 the action (\ref{GT27}), we identify
 \beq
  H=\int d^3x~\left[ {j^0 j^0 \over 2\mu^2}+{\mu^2\over
  2}(i\Tr(\s_3 g^{-1}\nabla g))^2\right]\label{CS7}
 \eeq
 We see that, unlike the case of the flavor skyrmions, we will need
 an ansatz for $j^0$ as well. For this purpose, consider the
 collective coordinate quantization of the soliton $g_s(\vx)$. Color
 transformations act on the field $g$ as $g\rightarrow Ug$. States with
 nonzero color charge can be generated from $g_s(\vx)$ by writing
 $g(\vx,t)=U(t) g_s(\vx)$. We also introduce spatial rotations $R(t)\in
 SO(3)$ and write a general ansatz as
 \beqar
  g(\vx,t)&=&U(t)~g_s(R\vx)\nonumber\\
  &=&U(t)S^{-1}(t)~g_s(\vx)~S(t)\label{CS8}
 \eeqar
 where $U(t)$, $R(t)$ (or its $(2\times 2)$-matrix version $S(t)$)
 represent the collective coordinates. Notice that while the ansatz
 (\ref{CS2}) is similar to what happens for flavor skyrmions,
 the collective coordinates enter into equation (\ref{CS8}) in a very
 different way. Using (\ref{CS8}) in the action, we find
 \beq
  S=-i\int dt~\bigl[ m~\Tr (\s_3U^{-1}\dot{U})+(n-m)~\Tr (\s_3S^{-1}\dot{S})\bigr]-\int
  dt ~H \label{CS9}
 \eeq
 where $m$, $n$ are given in terms of integrals involving
 combinations of $j^0$ and $\phi$. The action is given in terms of the
 KKS forms for $U$ and $S$ and so lead to representations of the
 color algebra ($SU(2)$ in this case) and the rotation algebra. For
 consistent quantization, with unitary representations of the two
 groups, we need $m$, $n$ to be integers. Our ansatz for $j^0$ must
 be consistent with this. Since there are two conditions we take an
 ansatz for $j^0$ which has two functions in it; the simplest choice
 is
 \beq
  j^0=f(r)+\left[ (v\cdot \hat{x})^2-{4\phi^{\prime 2}\over 3}\right] ~h(r)
  \label{CS10}
 \eeq
 where $v_i=i\Tr (\s_3g^{-1}\del_i g)$. Using the ansatz for $g$ we
 then find
 \beqar
  \int d^3x~ f(r)&=&n\nonumber\\
  {4\over 3}\int d^3x ~\left( f-{8~h~\phi^{\prime 2}\over 15}\right) \sin^2 \phi &=& n-m
  \label{CS11}
 \eeqar
 The energy functional can be worked out to be
 \beq
  \E={2\mu^2\over 3}\int d^3x ~\left(\phi^{\prime 2}+{2\sin^2 \phi\over
  r^2}\right)+{1\over 2\mu^2}\int d^3x ~\left( f^2+{64\over 45}(\phi^{\prime
  2}h)^2\right)\label{CS12}
 \eeq
 So far we have not used the specific form of $\phi$. We now simplify
 our ansatz further by taking $h(r)$ to be a constant and
 $f(r)= w (8\phi^{\prime 2}/ 15)$, where $w$ is another constant.
 Further taking $\phi$ to be given by the stereographic ansatz, the conditions
 (\ref{CS11}) are evaluated as
 \beq
  n={32\pi^2 R\over 15} w,~~~~ n-m={16\pi^2 R\over 15}(w-h)\label{CS13}
 \eeq
 which lead to
 \beq
  w={15\over 32\pi^2R}~n,~~~~ h={15\over 32\pi^2 R} (2m-n)\label{CS14}
 \eeq
 The energy function (\ref{CS12}) can also be evaluated easily as
 \beq
  \E (R)=8\pi^2\mu^2R+{5\over 16\pi^2\mu^2R^3}\left[(m-s)^2+{(m+s)^2\over 5}\right]
  \label{CS15}
 \eeq
 where $s=\half (n-m)$ is the spin of the soliton and $m$ defines the color
 charge. The collective coordinate action, upon quantization, gives an
 $(m+1)$-dimensional $SU(2)$ color multiplet and
 $(2s+1)$-dimensional representation of the rotation group. The
 minimum of $\E(R)$ occurs at
 \beq
  (\mu R)^4={15\over 128}{1\over \pi^4}\left[(m-s)^2+{(m+s)^2\over 5}\right]
  \label{CS16}
 \eeq
 and the variational estimate of the energy of the soliton is given as
 \beq
\E={8\mu\pi\over 3}(30)^{1/4}\left[(m-s)^2+{(m+s)^2\over 5}\right]^{1/4} \label{CS17}
\eeq
Going back to (\ref{CS8}), we see that the ansatz depends on $R(t)\in SO(3)$, so
only the integer spins can arise. This means that the the difference $n-m$ should be
an even integer. With this condition, formula (\ref{CS17}) gives the energy as a function of
the color charge and spin.

\section{Discussion}

In this paper we considered the general formulation of nonabelian fluid dynamics.
Since it is based on symmetry considerations and the generalization of the mechanics of
particles carrying nonabelian charges, there is a universality to this formulation.
It is then a very general result that the theory admits color skyrmion configurations.
We have obtained a (variational) mass formula for these as a function of spin and color charge, choosing a simple form for the Hamiltonian. Unlike the flavor skyrmions, there is no Wess-Zumino term, since there is no color anomaly. These
solitons are bosonic excitations of integral spin.

Clearly there are a number of interesting questions which need
further investigation. There is a hierarchy of energy scales which
is important in this context. The fluid approximation works in a
regime where there are few single-particle hard scattering events,
the collective effects are important, and before rehadronization
sets in. The form of the Hamiltonian has to be fixed by the equation
of state for the color degrees of freedom. (This is different from a
general equation of state, since there are other conserved charges
such as baryon number. Thus there are different partial pressures
and here we need the equation of state for the color partial
pressure as a function of the color charge density.) A more detailed
investigation of how our results depend on the form of the
Hamiltonian has to be carried out. Even after we have made the
simple choice (\ref{CS7}), the scale parameter $\mu$ is not
specified. While it can be related to the equation of state, we
expect that it is closely related to the magnetic screening in the
plasma. One reason for this is to notice that, if we absorb the
field $g$ as a gauge transformation of $A$, in a way analogous to
going to the unitary gauge in spontaneousy broken gauge theories,
our Hamiltonian has the form $H =\int \half \mu^2 A^3\cdot
A^3+\cdots$.

Another key issue is the production, detection and destruction of these solitons.
The conservation of the topological charge is vitiated only if the field $g(\vx, t)$ cannot be defined without singularities. So we expect that if such solitons are produced
in nuclear collisions, it should happen at the transition to the deconfined fluid phase.
Likewise, they should disappear due to rehadronization when the fluid description loses its validity. Dissipative effects can be introduced into our approach by modifying the equations of motion. While they can change many details, the qualitative features of the solitons should be unaltered.

We hope to address some of the
issues raised here in a future publication.
\vskip .2in\noindent
VPN thanks I. Zahed for a useful comment. This work
was supported in part by the National Science Foundation grant number
PHY-0244873 and by a CUNY Collaborative Research Incentive grant.

\end{document}